\documentclass{article}
\usepackage[dvips]{graphicx}
\begin{document}
\title{Shock Structures and Velocity Fluctuations in the Noisy 
Burgers and KdV-Burgers Equations}
\author{Hidetsugu Sakaguchi\\
Department of Applied Science for Electronics and Materials,\\ Interdisciplinary Graduate School of Engineering Sciences,\\
 Kyushu University, Kasuga, 816-8580, Japan}
\maketitle
{\bf abstract}\\
Statistical properties of the noisy Burgers and KdV-Burgers equations are numerically studied.  It is found that 
shock-like structures appear in the time-averaged patterns 
for the case of stepwise fixed boundary conditions. 
Our results show that the shock structure for the noisy KdV-Burgers equation has an oscillating tail, 
even for the time averaged pattern. Also, we find that 
the width of the shock and the intensity of the velocity fluctuations in the shock region increase with system size.
\\
\\
\\
\\
\section{Introduction}
The Burgers equation and the KdV-Burgers equation 
are well-known model equations used in the study of shock waves in fluids and plasmas.\cite{rf:1,rf:2} Spatial integration of the noisy Burgers equation yieldsthe Kardar-Parisi-Zhang equation.  
The KPZ equation has been intensively studied in the context 
of growing rough interfaces.\cite{rf:3} 
The Kuramoto-Sivashinsky equation is one of the simplest partial differential equations that exhibit spatiotemporal chaos. 
There is a conjecture that the statistical properties of the Kuramoto-Sivashinsky equation on large scales are closely related to those of the noisy Burgers equation.\cite{rf:4,rf:5} In previous studies, 
we have found shock structures in the time averaged patterns of the Kuramoto-Sivashinsky equation with fixed boundary conditions, where the boundary values are set as  $u(0)=-U_0,\,u(L)=U_0$.\cite{rf:6,rf:7} 
In this paper, we study some statistical properties of the noisy KdV-Burgers equation and the noisy Burgers equation with stepwise fixed boundary conditions. 

\section{Time-averaged shock structure and velocity fluctuations}
The noisy KdV-Burgers equation in one-dimension has the form
\begin{equation}
u_t=u_{xx}+d u_{xxx}+uu_x+\xi_x(x,t),
\end{equation}
where $u(x,t)$ is interpreted as a velocity variable on the interval $x\in [0,L]$, and the subscripts denote differentiation.   The noise $\xi(x,t)$ is assumed to be Gaussian white noise satisfying
\[\langle \xi(x,t)\rangle=0,\,\,\langle \xi(x,t)\xi(x,^{\prime},t^{\prime})\rangle=2D\delta(x-x^{\prime})\delta(t-t^{\prime}).\] 
If the parameter $d$ for the dispersion term in Eq.~(1) is equal to zero, this equation reduces to the noisy Burgers equation.  
The quantity $\int_0^L u(x)dx$ is conserved during the time evolution.

We consider firstly the case of periodic boundary conditions. 
The Fourier amplitude $u_{k_n}(t)=(1/\sqrt{L})\int_0^Lu(x)\exp(-ik_nx)dx$ ($k_n=2\pi n/L$) satisfies the Langevin equation
\begin{equation}
\frac{du_{k_n}}{dt}=-k_n^2u_{k_n}+ik_n\xi_{k_n}(t)+\{-idk_n^3u_{k_n}+\frac{1}{\sqrt{L}}\sum_m(ik_m)u_{k_n-k_m}u_{k_m}\},
\end{equation}
where $\xi_{k_n}(t)=(1/\sqrt{L})\int_0^L\xi(x)\exp(-ik_nx)dx$ is the Fourier amplitude for the noise $\xi(x,t)$, which satisfies $\langle \xi_{k_n}\rangle=0,\,\,\langle \xi_{k_n}(t)\xi_{k_m}(t^{\prime})\rangle=2D\delta_{n+m,0}\delta(t-t^{\prime})$.
Keeping only the first and second terms on the right-hand side of Eq.~(2), we obtain the simplest linear 
Langevin equation. Both the dispersion and the nonlinear terms in the brackets on the right-hand side in Eq.~(2) conserve the energy integral 
$\int_0^L|u(x)|^2dx=\sum_{n}|u_{k_n}|^2$. 
The probability distribution $P(\{u_{k_n}\})\propto\exp(-\sum_{n}|u_{k_n}|^2/2D)$ is therefore an equilibrium distribution for the corresponding Fokker-Planck equation,
\begin{equation}
\frac{\partial P}{\partial t}=\sum_{n}\frac{\partial }{\partial u_{k_n}}\left [-\left(-k_n^2u_{k_n}-idk_n^3u_{k_n}+\frac{1}{\sqrt{L}}\sum_{m}ik_mu_{k_n-k_m}u_{k_m}\right)P+2Dk_n^2\frac{\partial P}{\partial u_{k_{-n}}}\right ].
\end{equation}   
\begin{figure}[htb]
\begin{center}
\includegraphics[width=10cm]{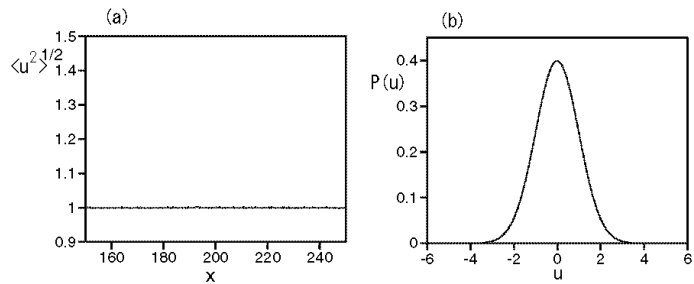}
\end{center}
\caption{(a) Time-averaged profile of $\langle u^2\rangle^{1/2}$ for the noisy KdV-Burgers equation with $d=2$ under periodic boundary conditions. (b) The velocity distribution $P(u)$ at $x=L/2$.}
\label{fig:1}
\end{figure}
\begin{figure}[htb]
\begin{center}
\includegraphics[width=12cm]{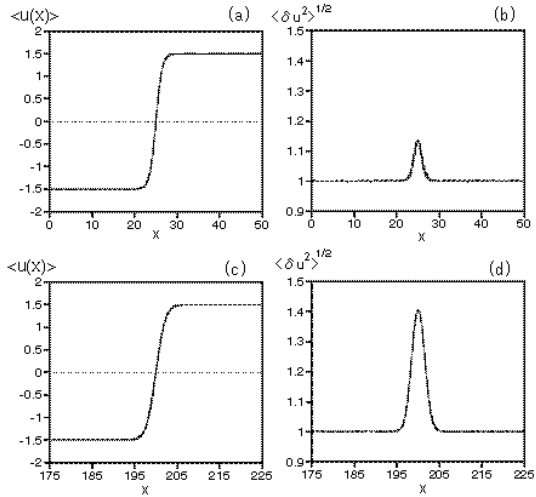}
\end{center}
\caption{(a) Time-averaged pattern of $u(x)$ (solid curve) for the noisy Burgers equation under the stepwise fixed boundary conditions $u(0)=-1.5$ and $u(L)=1.5$ with $L=50$.  The dashed curve represents  $1.5\tanh(0.71(x-L/2))$. The two curves are almost indiscernable.
(b) Time-averaged profile of $\langle \delta u^2\rangle^{1/2}=(\langle u^2(x)\rangle-\langle u(x)\rangle^2)^{1/2}$ for $L=50$. The dashed curve is $1+0.14/\cosh^2\{0.75(x-L/2)\}$ (c) Time averaged pattern of $u(x)$ (solid curve)  and $1.5\tanh(0.475(x-L/2))$ (dashed curve) for $L=400$.
(d) Time averaged profile of $\langle \delta u^2\rangle^{1/2}=(\langle u^2(x)\rangle-\langle u(x)\rangle^2)^{1/2}$ for $L=400$.}
\label{fig:2}
\end{figure}
\begin{figure}[htb]
\begin{center}
\includegraphics[width=11cm]{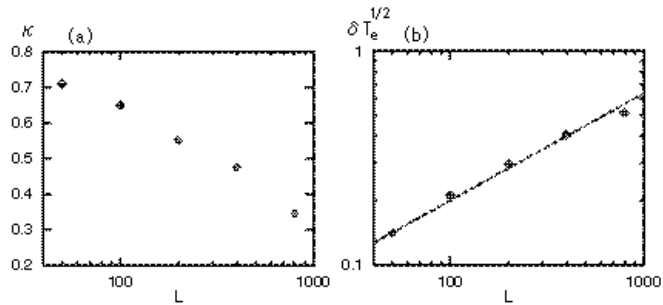}
\end{center}
\caption{(a) System size dependence of $\kappa$, where the time averaged shock structure is approximated as $1.5\tanh\{\kappa(x-L/2)\}$. 
(b) System size dependence of the excess part $\delta T_e^{1/2}$ of the standard deviation of the velocity distribution. The dashed line corresponds to a line of $L^{1/2}$.}
\label{fig:3}
\end{figure}
\begin{figure}[htb]
\begin{center}
\includegraphics[width=10cm]{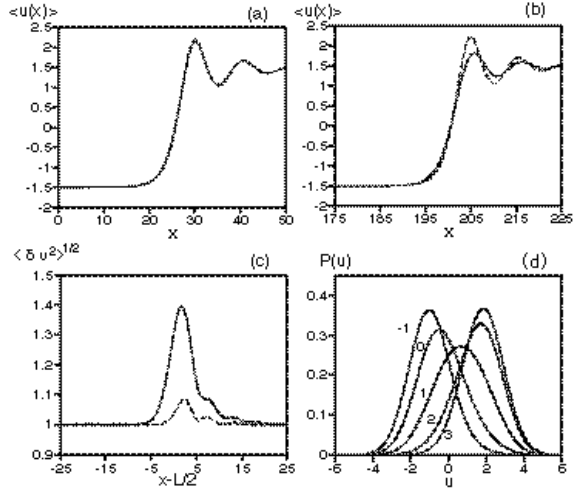}
\end{center}
\caption{(a) Time averaged-pattern of $u(x)$ for the noisy KdV-Burgers equation with $d=2$ under the stepwise fixed boundary conditions $u(0)=-1.5$ and $u(L)=1.5$ with $L=50$. The dashed curve is the stationary solution for the KdV-Burgers equation with $d=2$ without the noise term.
(b) Time-averaged pattern of $u(x)$ for the noisy KdV-Burgers equation with $L=400$. The dashed curve is the stationary solution for the deterministic KdV-Burgers equation.
(c) Time-averaged profile of $\langle \delta u^2\rangle^{1/2}=(\langle u^2(x)\rangle-\langle u(x)\rangle^2)^{1/2}$ around the shock positions for $L=50$ (dashed curve) and $L=400$ (solid curve). (d) The velocity distribution $P(u)$ at $x=L/2-1,\,L/2,\,L/2+1,\,L/2+2$ and $L/2+3$ for $L=400$. The numbers in the figure denote the distance, $x-L/2$, from the center.}
\label{fig:4}
\end{figure}
\begin{figure}[htb]
\begin{center}
\includegraphics[width=10cm]{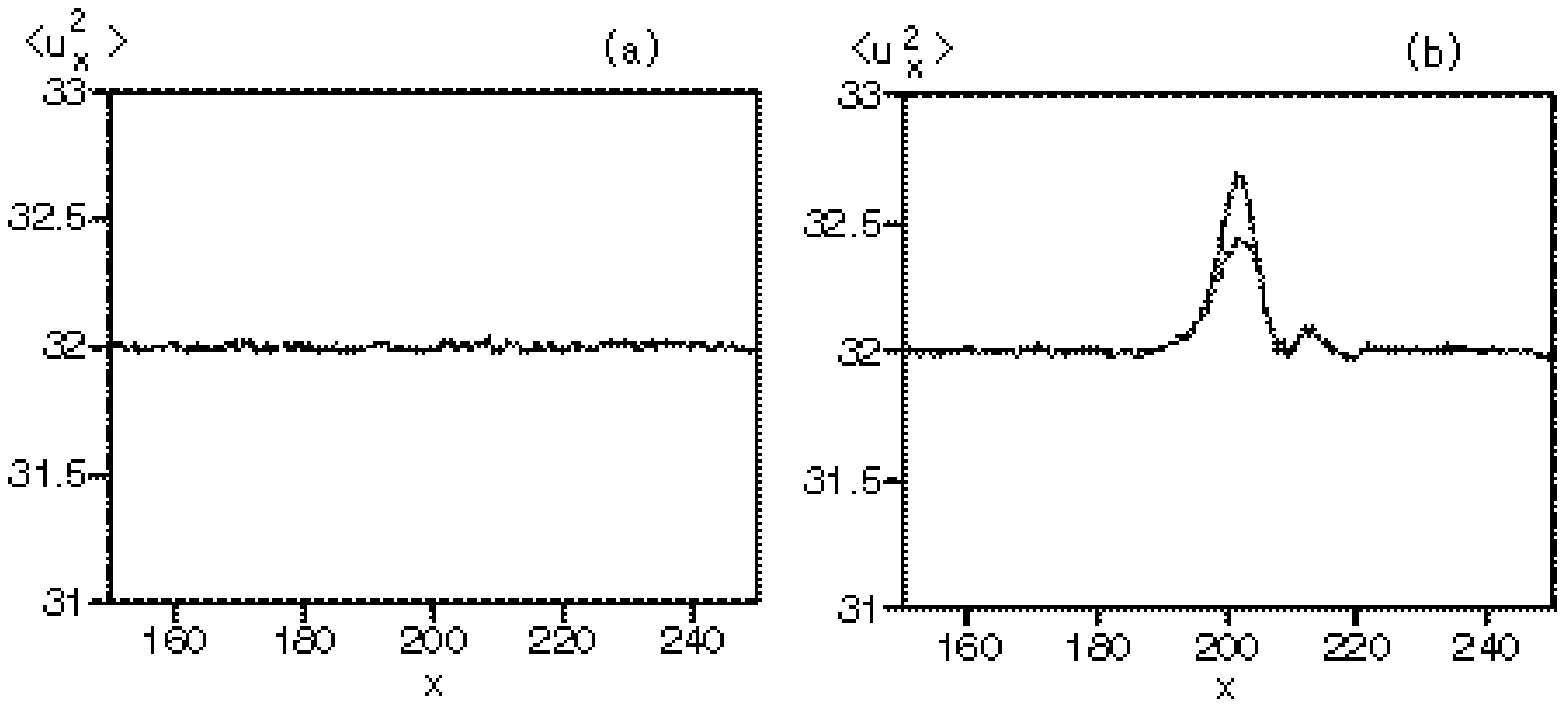}
\end{center}
\caption{(a) Energy dissipation rate $\langle u_x^2\rangle$ 
 for the KdV-Burgers equation with $d=2$ and system size $L=400$ under periodic boundary conditions. (b) Energy dissipation rate $\langle u_x^2\rangle$ (solid curve) and the contribution from the velocity fluctuations $\langle (\delta u)_x^2\rangle$
 (dashed curve) for the KdV-Burgers equation with $d=2$ under the stepwise fixed boundary conditions $u(0)=-1.5$ and $u(L)=1.5$ for $L=400$.}
\label{fig:5}
\end{figure}
In real space, the equilibrium distribution can be rewritten $P(\{u(x)\})\propto\exp(-\int_0^Lu(x)^2dx/2D)$.  
We have confirmed this thermal equilibrium distribution in numerical simulation, using the Heun method (the second order Runge- Kutta  method) with temporal and spatial step sizes  $\Delta t=0.00025$ and $\Delta x=1/4$. The parameter values are chosen to be $L=400, d=2$ and $D=1/4$. The numerical simulation was performed up to time $t_f=101250$,  and the statistical averages were calculated as the long time averages between $t=1250$ and $t=101250$. The equilibrium distribution for the spatially discretized system is written  $P(\{u_i\})\propto \exp(-\sum_iu_i^2\Delta x/2D)$, where $u_i$ represents $u(x)$ at $x=i\Delta x$.  
The average value of $u(x)$ is zero, and therefore the variance of the velocity   is $\langle u(x)^2\rangle=D/\Delta x$ for any $x$. Figure 1(a) displays the profile of $\langle u^2(x)\rangle^{1/2}$.  The average value of $u^2$ is almost 1 for any $x$. Figure 1(b) displays the velocity distribution $P(u(L/2))$ at $x=L/2$ and it is compared with the Gaussian distribution with variance 1. No difference can be discerned in this plot. The thermal equilibrium distribution was also obtained  for the noisy Burgers equation with $d=0$. 

Next, we performed numerical simulations with stepwise fixed boundary conditions.
The boundary values we used are $u(0)=-U_0$ and $u(L)=U_0$ for $L=50$.
The initial condition is $u(x)=U_0\tanh(U_0(x-L/2)/2)$.
During the simulation, at regular intervals,  we confirmed that the condition $\int_0^Ludx=0$ was satisfied. 
Figure 2 displays the numerical results for the noisy Burgers 
equation with $U_0=1.5$.
The solid curve in Fig.~2(a) represents the time-averaged profile $\langle u(x)\rangle$, and it is compared with the curve of $1.5\tanh 0.71(x-L/2)$. A shock structure appears 
in the time-averaged pattern. 
We have checked that the time-averaged pattern depends only weakly on the total time $t_f$.
If the noise term is absent, i.e., $D=0$, the shock solution to the Burgers equation is $u(x)=U_0\tanh\{U_0/2(x-L/2)\}$. The width of the time-averaged shock structure  in Fig.~2(a) is almost the same as that of the deterministic equation.   Figure 2(b) displays a profile of the standard deviation $\langle \delta u^2\rangle^{1/2}=\langle (u(x)-\langle u(x)\rangle)^2\rangle^{1/2}$. In the flat region,  distant from the shock region, the standard deviation of the velocity distribution is almost 1, but, it increases in the shock region. 
The variance of the velocity distribution is interpreted as an effective temperature.   
If the position of the shock  is assumed to fluctuate due to the noise term, the profile of $u$ would fluctuate as $u(x)\sim U_0\tanh\{U_0/2(x-L/2-\delta x)\}\sim U_0\tanh\{U_0/2(x-L/2)\}-U_0^2\delta x/[2\cosh^2\{U_0/2(x-L/2)\}]$. Then, the deviation of $\langle \delta u^2\rangle^{1/2}$ from 1 can be approximated as $\{U_0^2/(2\cosh^2(U_0/2(x-L/2))\}\langle \delta x^2\rangle^{1/2}$. The numerical result for the standard deviation of the velocity distribution is fit well by the form  $1+0.14/\cosh^2\{0.75(x-L/2)\}$, as shown in Fig.~2(b). The slight deviation of the time averaged profile from $U_0\tanh\{U_0/2(x-L/2)\}$ may also be interpreted as a blurring effect resulting from the fluctuations of the shock position.
(This type of deviation of the time-averaged profile from the deterministic shock is observed for sharp shocks, for which $U_0$ is sufficiently large. Such a deviation was not clearly observed in a previous investigation in the case of  smooth shocks.\cite{rf:7}) 

The profile of the time-averaged pattern and the standard deviation does not depend on the the total simulation time $t_f$. However, it depends strongly on the system size, $L$.
Figure 2(c) displays the time-averaged profile $\langle u(x)\rangle$ for $L=400$, and it is approximated as $1.5\tanh 0.475(x-L/2)$. The width of the shock structure is larger for this larger size system. Figure 2(d) displays the standard deviation of the velocity distribution for $L=400$. We see that the increase of the effective temperature in the shock region is larger than in the case of $L=50$. 
The time-averaged shock structure can be approximated as $1.5\tanh\{\kappa(x-L/2)\}$, where $\kappa$ is the inverse of the shock width.  
The system size dependence of $\kappa$ and the excess part of the standard deviation of the velocity distribution $\delta T_e^{1/2}=\langle \delta u^2\rangle^{1/2}-1$, at $x=L/2$ are displayed in Fig.~3. The quantity  
seems to increase as $\delta T_e\sim L^{1/2}$.  We do not yet understand the system size dependence well for this noisy Burgers equation.  A simple interpretation is as follows: the value of $u$ fluctuates randomly around $\pm U_0$ in the region distant from the shock position, therefore, the integration $\int u(x)dx$  between 0 and $L$ except for the shock region takes a random value of  $O(L^{1/2})$. The total value of $\int_0^Lu(x)dx$ is conserved to be 0 in our model.  To compensate the total fluctuations of $u(x)$ in the surrounding region, the position of the shock moves necessarily by $-O(L^{1/2})$. Shock fluctuations have been studied in an asymmetric simple-exclusion process with a blockage site, which is one of the simplest of the driven diffusive lattice gas madels. In that model, the amplitude of the fluctuations of the shock position increases as $L^{1/2}$ or $L^{1/3}$, depending on the system parameters.\cite{rf:8}  This behavior may be related to our results. (In contrast, for an asymmetric simple-exclusion process with open boundaries, the shock position exhibits a random walk.\cite{rf:9})    
 
Figures 4(a) and (b) display the time-averaged profiles in the cases $L=50$ and $400$ for the noisy KdV-Burgers equation with $d=2$. We see that the time-averaged  profile exhibits damping oscillations for $x>L/2$, and 
it is asymmetric about the center of the shock structure. 
This is characteristic of the KdV-Burgers equation with a dispersion term. 
 The dashed curve is the numerically obtained shock solution for the KdV-Burgers equation (1) without the noise term for $d=2$.
It is seen that this curve nearly coincides with the time-averaged structure for $L=50$. However,  
the width of the  time-averaged shock structure is longer for $L=400$, and the fit to the deterministic shock is worse for the larger system. 
This may also be due to the fluctuations of the shock positions.
Figure 4(c) displays the standard deviation of the velocity distribution $\langle \delta u^2\rangle^{1/2}$ for $L=50$ (dashed curve) and 400 (solid curve). It is seen that here too, the effective temperature increases in the shock region. The profile of the effective temperature is also asymmetric around the shock center. 
The increase of the effective temperature for $L=400$ is clearly larger than for $L=50$.  Figure 4(d) displays the velocity distribution $P(u(x))$ at $x=L/2-1,L/2,L/2+1,L/2+2$ and $L/2+3$ for $L=400$.  The velocity distribution approaches  Gaussian distribution with variance 1 as $|x-L/2|$ is increased. 
The variance of the velocity distribution is larger than 1 and the distribution  becomes asymmetric near the shock region. 

\section{Energy dissipation rate}
The nonequilibrium states discussed above appear as a result of the  stepwise fixed boundary conditions. Energy is injected into the system through the boundary conditions, and energy dissipation occurs in the region of the shock.   
From Eq.~(1), the time evolution of the energy density can be written as
\begin{equation}
\frac{1}{2}\frac{\partial u^2}{\partial t}=\frac{\partial (uu_x)}{\partial x}+d \left[\frac{\partial (uu_{xx})}{\partial x}-\frac{1}{2}\frac{\partial u_x^2}{\partial x}\right ]+\frac{1}{3}\frac{\partial u^3}{\partial x}-u_x^2+u\xi_x.
\end{equation}
If $D=0$ and $U_0\ne 0$, the third term on the right-hand side is related to the energy injection at the boundaries, and the fourth term represents the energy dissipation.  The energy injection from the boundary conditions is $2/3U_0^3$. The 
total energy dissipation is expressed as $Q=\int_0^Lu_x^2dx$, and it is equal to $2/3U_0^3$. (This can be explicitly shown by the integration of $u_x^2=(\partial_x\{U_0\tanh U_0(x-L/2)/2\})^2$ for the case  $d=0$.) If $D\ne 0$ and  periodic boundary conditions are employed, the total energy dissipation rate, $Q=\int_0^Lu_x^2dx$, is 
equal to the quantity $Q^{\prime}=\int_0^Lu\xi_xdx$, which is interpreted as 
the heat flow from the heat reservoir expressed by the noise term.
The total entropy production is therefore zero in the thermal equilibrium state. 
We have calculated the time average $\langle u_x^2(x)\rangle$ 
in our simulation model with $\Delta x=1/4$ and $d=2$. 
Figure 5(a) displays the profile of   $\langle u_x^2(x)\rangle$ 
for the case of periodic boundary conditions. In this case, the energy dissipation rate is nearly 32. Also, it can be shown that the average $\langle u(x,t)\xi_x\rangle=2D/(\Delta x)^3=32$ in the spatially discretized system with $D=1/4$ and $\Delta x=1/4$.  
That is, we have found numerically that the average of the entropy production rate is zero for any $x$. The probability distribution of $u_x$ is well approximated as a Gaussian distribution. The solid curve in Fig~5(b) displays the profile of $\langle u_x^2(x)\rangle$ for the noisy KdV-Burgers equation in the case of $d=2$ with the fixed boundary conditions $u(0)=-1.5,u(L)=1.5$ for $L=400$. Note that the energy dissipation rate is larger than the equilibrium value 32 in the shock region.  The energy dissipation term $u_x^2(x)$ is written as a sum of two components: $u_x^2(x)=\langle u(x)\rangle_x^2+(\delta u(x))_x^2$. The dashed curve in Fig.~5(b) represents the contribution by $\langle (\delta u)_x^2\rangle$ from the velocity fluctuations.  The rate of energy dissipation resulting from the velocity fluctuations seems to be larger than that resulting from  the time-averaged pattern in this simulation. We calculated the energy dissipation rate resulting from the velocity fluctuations also for $L=50$. Clear difference by the system size was not found in this energy dissipation rate resulting from the velocity fluctuations in contrast to the velocity fluctuations shown in Fig.~4(c).

\section{Summary and discussion}
We have performed numerical simulation of the noisy Burgers and KdV-Burgers equations. We have found that a thermal equilibrium state is realized for periodic  boundary conditions and a shock structure appears as a nonequilibrium state for  stepwise fixed boundary conditions. 
The statistical properties of nonequilibrium states have been studied by many researchers. In particular, the 
statistical properties of nonequilibrium liquids have been intensively studied using molecular dynamics simulations.\cite{rf:10}  
Nonequilibrium states, including shocks, have been studied using driven lattice gas models, most notably those with asymmetric exclusion processes.\cite{rf:11,rf:8} 
We have studied the statistical properties of the 
nonequilibrium states created by the boundary conditions with our Langevin-type models.  
Increases in both the effective temperature and the energy dissipation rate caused by the velocity fluctuations are observed in the shock region. 
This increase of the effective temperature may be due to the fluctuations of the shock position. The intensity of the fluctuations depends strongly on the system size. The increases of the energy dissipation rate resulting from the velocity fluctuation does not so depend on the system size. The mechanisms of the increases of the two statistical quantities in the shock region may be different.
 The time-averaged shock structure in our model may be closely related to the macroscopic shock structure found in the asymmetric simple exclusion model. However, a detailed investigation of their relation is left to a future study. 


\begin{thebibliography}{99}
\bibitem{rf:1} V.~I.~Karpman, {\it Nonlinear Waves in Dispersive Media} (Pergamaon Press, 1975).
\bibitem{rf:2} R.~Z.~Sagdeev, D.~A.~Usikov and G.~M.~Zaslavsky, {\it Nonlinear Physics: From Pendulum to Turbulence and Chaos} (Harwood Academic Publishes, Chur,1988).
\bibitem{rf:3} M.~Kardar, G.~Parisi and Y.~C.~Zhang, Phys. Rev. Lett. {\bf 56} (1986), 889.
\bibitem{rf:4} V.~Yakhot, Phys. Rev. A {\bf 24} (1981), 642.
\bibitem{rf:5} K.~Sneppen, J.~Krug, M.~H.~Jensen, C.~Jayaprakash and T.~Bohr, Phys. Rev. A {\bf 46} (1992), R7351.
\bibitem{rf:6} H.~Sakaguchi, Phys. Rev. E {\bf 62} (2000), 8817.
\bibitem{rf:7} H.~Sakaguchi, Prog. Theor. Phys. {\bf 107} (2002), 879.
\bibitem{rf:8} S.~A.~Janowsky and J.~L.~Lebowitz, Phys. Rev. A {\bf 45} (1992), 618.
\bibitem{rf:9} A.~B.~Kolomeisky, G.~M.~Sch\"utz, E.~B.~Kolomeisky and J.~P.~Straley, J. of Phys. A {\bf 31} (1998), 6911. 
\bibitem{rf:10} D.~J. Evans and G.~P.~Morris, {\it Statistical Mechanics of Nonequilibrium Liquids} (Academic Press, London, 1990). 
\bibitem{rf:11} J.~Lebowitz, E.~Presutti and H.~Spohn, J. Stat. Phys. {\bf 51} (1988), 841.
\end{thebibliography}
\end{document}